\documentclass[aps,prl,reprint,amsmath,showpacs,superscriptaddress]{revtex4-1}
\usepackage{graphicx}
\usepackage{textcomp}
\usepackage{float}
\usepackage{xcolor}
\usepackage[normalem]{ulem}

\begin{document}

\title{Excellent electronic transport in  heterostructures of graphene and monoisotopic boron-nitride grown at atmospheric pressure}

\author{J.~Sonntag}
\email{Corresponding author: sonntag@physik.rwth-aachen.de}
    \affiliation{JARA-FIT and 2nd Institute of Physics, RWTH Aachen University, 52074 Aachen, Germany}
    \affiliation{Peter Gr{\"u}nberg Institute (PGI-9), Forschungszentrum J{\"u}lich, 52425 J{\"u}lich, Germany}

\author{J.~Li}
    \affiliation{Tim Taylor Department of Chemical Engineering, Kansas State University, Manhattan, Kansas 66506, United States}
    
\author{A.~Plaud}
    \affiliation{Groupe d{'}Etude de la Mati\`{e}re Condensée (GEMaC), Universit\'{e} de Versailles St Quentin en Yvelines, CNRS, Universit\'{e} Paris Saclay, 78035 Versailles, France}
    \affiliation{Laboratoire d{'}Etude des Microstructures (LEM), ONERA, CNRS, Universit\'{e} Paris-Saclay, 92322 Ch\^{a}tillon, France}
    
\author{A.~Loiseau}
    \affiliation{Laboratoire d{'}Etude des Microstructures (LEM), ONERA, CNRS, Universit\'{e} Paris-Saclay, 92322 Ch\^{a}tillon, France}
    
\author{J.~Barjon}
    \affiliation{Groupe d{'}Etude de la Mati\`{e}re Condensée (GEMaC), Universit\'{e} de Versailles St Quentin en Yvelines, CNRS, Universit\'{e} Paris Saclay, 78035 Versailles, France}

\author{J.~H. Edgar}
    \affiliation{Tim Taylor Department of Chemical Engineering, Kansas State University, Manhattan, Kansas 66506, United States}

\author{C.~Stampfer}
    \affiliation{JARA-FIT and 2nd Institute of Physics, RWTH Aachen University, 52074 Aachen, Germany}
    \affiliation{Peter Gr{\"u}nberg Institute (PGI-9), Forschungszentrum J{\"u}lich, 52425 J{\"u}lich, Germany}

\date{\today}

\begin{abstract}
    Hexagonal boron nitride (BN), one of the very few layered insulators, plays a crucial role in 2D materials research.
	In particular, BN grown with a high pressure technique has proven to be an excellent substrate material for graphene and related 2D materials, but at the same time very hard to replace.
	Here we report on a method of growth at atmospheric pressure as a true alternative for producing BN for high quality graphene/BN heterostructures.
	The process is not only more scalable, but also allows to grow isotopically purified BN crystals.
    We employ Raman spectroscopy, cathodoluminescence, and electronic transport measurements to show the high-quality of such monoisotopic BN and its potential for graphene-based heterostructures.
	The excellent electronic performance of our heterostructures is demonstrated by well developed fractional quantum Hall states, ballistic transport over distances around $10\,\mathrm{\mu m}$ at low temperatures and electron-phonon scattering limited transport at room temperature.
\end{abstract}

\maketitle

In recent years, a large number of two-dimensional (2D) materials have been discovered~\cite{Ajayan2016Aug,Geim2013Jul,Novoselov2016Jul,Butler2013Apr,Mounet2018Feb,Haastrup2018Sep}, investigated and used in first prototype devices. 
These materials cover almost all types of different material classes, including  metals, semimetals, semiconductors, insulators, superconductors~\cite{Xi2015Jul,Cao2015Aug}, and even ferromagnets~\cite{Bonilla2018Feb,Huang2017Jun}.
However, the number of insulators is very limited since up to now only hexagonal boron nitride (BN) is available as a true 2D layered insulator.
This gives hexagonal BN a special significance in particular since most properties of functional 2D materials - which consist only of surface atoms -  are strongly influenced by the direct environment, making substrate materials and capping layers highly crucial for the effective material quality and device performance.  
Indeed, encapsulating 2D materials in hexagonal BN opened the way for improving device performance~\cite{Dean2010,Wang2013,Ajayan2016Aug,Geim2013Jul,Novoselov2016Jul,Dauber2015May} and to thoroughly studying the rich electronic, mechanical and optical properties of 2D materials.  
For example, phenomena observed in 2D materials thanks to the BN encapsulation range from the fractional quantum Hall effect~\cite{Dean2011,Kim2019Nov,Zibrov2018Jul}, the Hofstadter's butterfly in moir\'{e} superlattices~\cite{Dean2013}, charge density waves~\cite{Xi2015}, single photon emitters~\cite{Koperski2015May} and to superconductivity in twisted bilayer graphene~\cite{Cao2018}.
All these observations crucially rely on the material quality of the encapsulating BN.
So far the very best and most used hexagonal BN is produced via a high pressure  high temperature (HPHT) process ($>30\,\mathrm{kbar}$)~\cite{Watanabe2004}, and is kindly provided by T. Taniguchi and K. Watanabe to the extending 2D materials research community~\cite{Zastrow2019Aug}. 
While the HPHT method produces BN of the highest quality, it is clearly limited in its scalability. 
An alternative method of growth using atmospheric pressure and high temperature (APHT)~\cite{Kubota2007Aug,Kubota2008Mar,Hoffman2014May,Liu2017} is not only more suitable for larger scales~\cite{Liu2017,Kumaravadivel2019Jul}, but furthermore allows for the easy control of isotopic concentrations~\cite{Liu2018}.

\begin{figure*}[tb]
	\centering
	\includegraphics{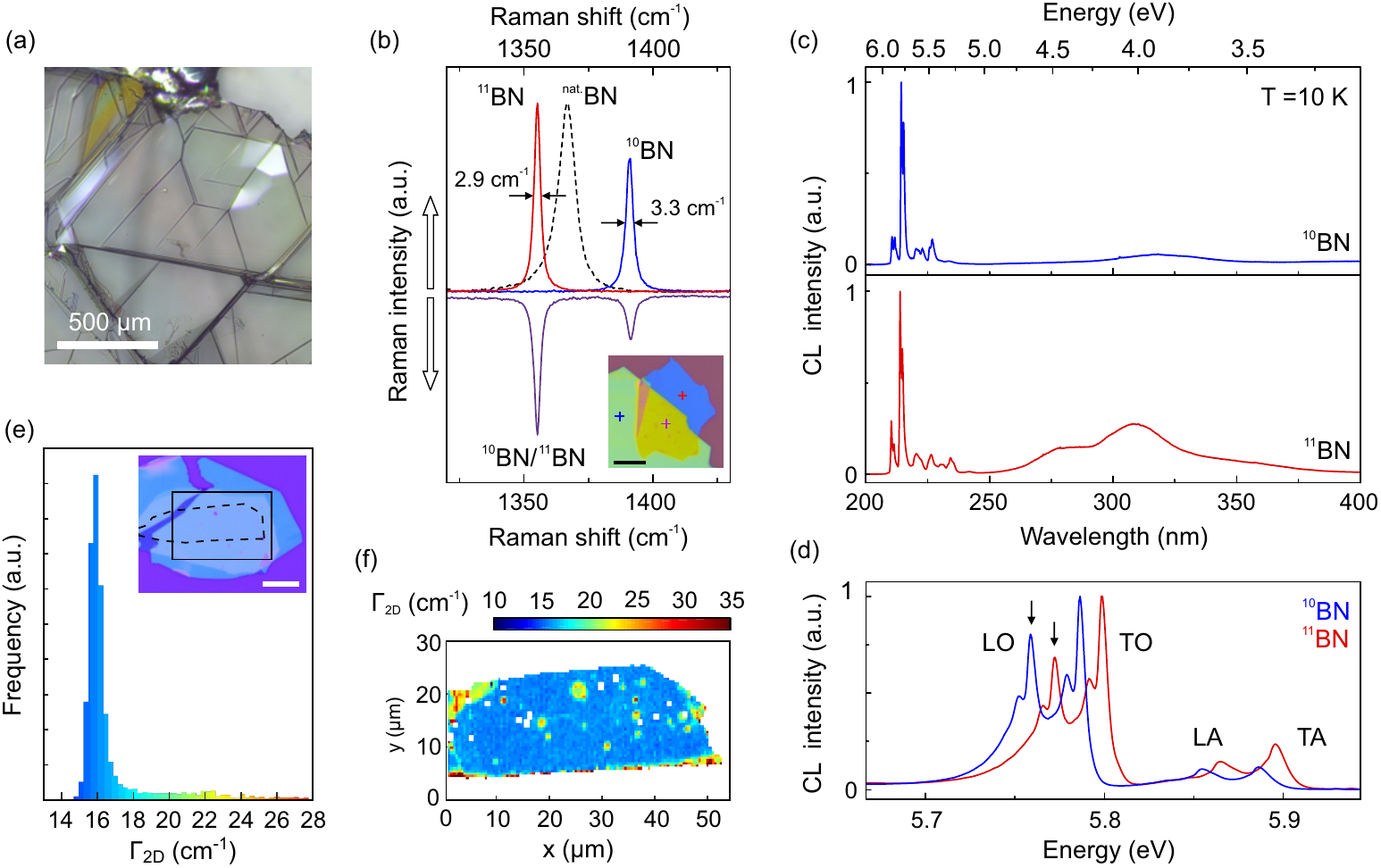}
	\caption{(a) Micrograph of a \textsuperscript{11}BN crystal.
	(b) Raman spectra of a heterostructure made from \textsuperscript{10}BN and \textsuperscript{11}BN, recorded at the positions displayed within the inset. Scale bar: $10\,\mathrm{\mu m}$.
	The black dashed line represents the Raman response from BN with natural isotope concentrations grown via a APHT process.
	(c) Cathodoluminescence spectra for both \textsuperscript{10}BN and \textsuperscript{11}BN crystals, taken at $T=10\,\mathrm{K}$ with an acceleration voltage of $5\,\mathrm{kV}$.
	(d) High resolution spectra of the phonon assisted exciton recombinations.
	(e) Distribution of the Raman 2D-peak width within a \textsuperscript{10}BN/graphene heterostructure with a maximum at $\Gamma_\mathrm{2D}<16\,\mathrm{cm^{-1}}$.
	Inset: Optical image of a BN/graphene/BN heterostructure made with monoisotopic BN.
	The dashed lines indicate the graphene flake. Scale bar: $20\,\mathrm{\mu m}$.
	(f) Spatially resolved Raman map of the 2D-peak width $\Gamma_\mathrm{2D}$.
	}
	\label{fig:Fig1}
\end{figure*}

In comparison to BN with the natural distribution of boron ($19.9\,\%$~\textsuperscript{10}B and $80.1\,\%$~\textsuperscript{11}B), isotopically purified BN exhibits a decrease in phonon-phonon scattering, leading to enhanced thermal transport~\cite{Morelli2002Nov,Chang2006Aug,Lindsay2011Oct} and to ultralow-loss polaritons~\cite{Giles2017Dec}, making monoisotopic BN interesting for nanophotonics.
As the effective thermal management is one of the bottlenecks in nanoelectronics, monoisotopic BN is further a promising building block for high performance and high-power electronic applications, especially since the cooling can be further enhanced by hyperbolic phonon polariton cooling~\cite{Tielrooij2017Nov,Yang2017Nov,Baudin2019Sep}. 
Furthermore, isotope purification modifies the electron density distribution and the van der Waals interaction between layers~\cite{Vuong2017Dec}, which could alter the characteristics of the heterostructures, especially in regards to modifications to the phonon band structure with possible impact on the electronic transport~\cite{Banszerus2019Sep}.  
Thus the APHT grown monoisotopic BN is not only interesting for applications, but also for fundamental research.
However, so far there is not much known about the quality of this material with respect to its use for electronic devices based on heterostructures of graphene and such BN.

Here we report on electronic transport measurements on heterostructures based on graphene encapsulated in APHT grown monoisotopic BN~\cite{Liu2017,Liu2018}.
We assess the quality of the APHT grown monoisotopic BN via confocal Raman and cathodoluminescence spectroscopy measurements and we show that the graphene heterostructures -- based on both \textsuperscript{10}BN and \textsuperscript{11}BN -- exhibit excellent electronic performance, absolutely equivalent to state-of-the-art heterostructures based on HPHT hexagonal BN.
At low temperature the charge carrier mobility is only limited by the device size, fractional quantum Hall effect appears below $14\,\mathrm{T}$ and magnetic focusing experiments allow to extract mean free paths very close to 10~$\mathrm{\mu}$m. 
At room temperature the carrier mobility of all fabricated devices is only limited by electron-phonon scattering, even possibly outperforming present BN/graphene devices, making this material highly interesting for applications that require ultra-high carrier mobilities.  

The monoisotopic hexagonal boron nitride crystals were grown with the same metal flux method as reported in Ref.~\cite{Liu2018}.
In short, elemental boron powder with high isotope purity ($99.22\,\%$ \textsuperscript{10}B or $99.41\,\%$ \textsuperscript{11}B) are mixed with nickel and chromium powder and placed into an alumina crucible within an alumina tube furnace.
During the growth process at 1550~$^{\circ}$C, a continuous flow of N\textsubscript{2} and H\textsubscript{2} gases at a constant pressure of 850 Torr through the furnace is applied.
Note that all the nitrogen elements in the BN crystal originate from the flowing N\textsubscript{2} gas and that the natural distribution of nitrogen isotopes already consists of $>99.6\,\%$~\textsuperscript{14}N~\cite{Weast1984}, creating an isotopically pure BN crystal.
During the cool down (0.5$^\circ$C/h) of the furnace the BN crystals precipitate on the metal surface.
After removing the crystals from the metal surface (see Fig.~\ref{fig:Fig1}), we prepare thin flakes by exfoliating the crystals onto 285~nm SiO\textsubscript{2} on silicon.

\begin{figure}[tb]
	\centering
	\includegraphics{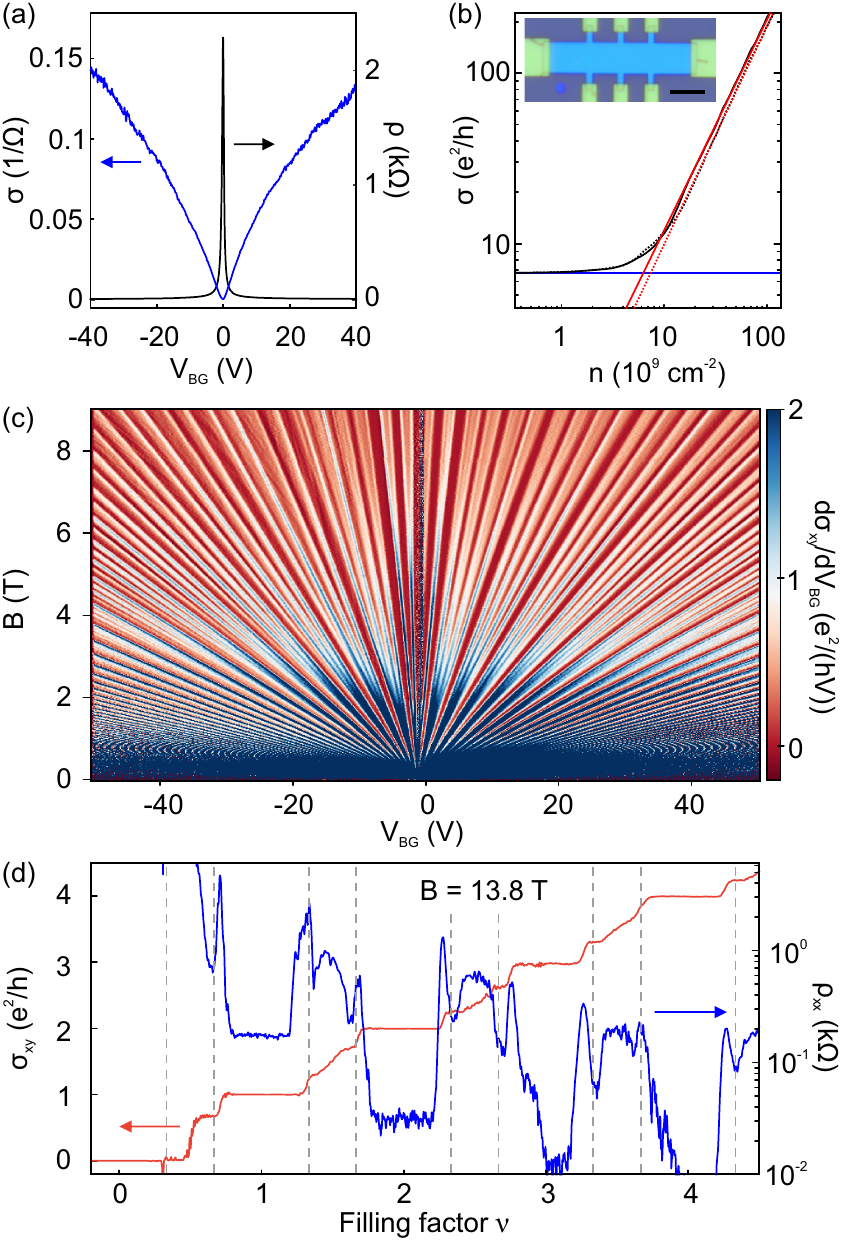}
	\caption{Electrical transport on heterostructures built with isotopically cleaned BN. (a) Electrical conductivity~$\sigma$ and resistivity~$\rho$ of a \textsuperscript{10}BN heterostructure as a function of applied backgate voltage $V_{BG}$ at $T=1.7\,\mathrm{K}$.
		(b) Double logarithmic plot $\sigma$ vs charge carrier density $n$ for determining the charge carrier inhomogeneity $n^*$. The dashed lines indicate hole doping. Inset: optical image of a finished device. Scale bar: $5\,\mathrm{\mu m}$.
		(c) Landau fan showing differential conductance $d\sigma/dV_{BG}$ as a function of $V_{BG}$ and B. Landau level begin to form below 0.5~T and the Landau level degeneracy starts to lift already at approx. 2~T.
		(d) Quantized Hall voltage $\sigma_{xy}$ (red) and longitudinal resistance $\rho_{xx}$ (blue) as a function of filling factor. Fractional filling factors $\nu=1/3, 2/3, 4/3...$ are indicated by dashed, gray lines.
	}
	\label{fig:Fig2}
\end{figure}
    
We build heterostructures based on monoisotopic BN via dry-stacking of the individual flakes~\cite{Wang2013}.
A heterostructure built from different isotopes \textsuperscript{10}BN and \textsuperscript{11}BN is shown as inset in Fig.~\ref{fig:Fig1}(b).
We characterize the heterostructures (both graphene and BN) via confocal Raman spectroscopy~\cite{Graf2007Feb,Forster2013Aug}.
As seen in Fig.~\ref{fig:Fig1}(b) we can clearly distinguish the in-plane $E_\mathrm{2g}$ Raman mode of the two isotopes positioned at $1391.0\,\mathrm{cm^{-1}}$ and $1355.2\,\mathrm{cm^{-1}}$ for \textsuperscript{10}BN and \textsuperscript{11}BN respectively.
The width of this peak is commonly used as a measure for the crystallinity of BN~\cite{Schue2016}. 
In the highest quality HPHT grown BN crystals with the natural distribution of boron the lowest width observed is $7.3\,\mathrm{cm^{-1}}$~\cite{Schue2016}, mainly broadened by isotopic disorder.
Similarly, we observe a comparable width of $7.5\,\mathrm{cm^{-1}}$ for natural BN grown with a APHT process (see dashed line in Fig.~1(b)). 
For the monoisotopic BN we observe values of $\Gamma_\mathrm{\textsuperscript{10}BN}=3.3\,\mathrm{cm^{-1}}$ and $\Gamma_\mathrm{\textsuperscript{11}BN}=2.9\,\mathrm{cm^{-1}}$. 
This narrow width of the E\textsubscript{2g} mode indicates a high isotope purity and a high crystal quality.

For the characterization of high quality hBN crystals, information gained from Raman spectroscopy is limited and needs to be completed by luminescence experiments~\cite{Schue2016}.
To this end, we record cathodoluminescence (CL) spectra at low temperature ($T=10\,\mathrm{K}$) with an acceleration voltage of $5\,\mathrm{kV}$ and correct them for the spectral response of the detection system~\cite{Schue2019Feb}, see Fig.~\ref{fig:Fig1}(c). 
The radiative recombination of free excitons dominates the CL spectra of both \textsuperscript{10}BN and \textsuperscript{11}BN exfoliated crystals with a maximum at 215 nm.
The intensity of luminescence signals related to structural defects (227 nm) and deep defects (broadbands near 300 nm) remains much weaker than exciton signal (17.8 and 3.6 times lower for \textsuperscript{10}BN and \textsuperscript{11}BN),  which indicates that the quality of APHT BN crystals is close to HPHT ones~\cite{Schue2016}.
The control of the boron isotopes results in a slight energy shift for the series of narrow lines due to various phonon modes, including longitudinal and transverse optical (LO, TO) and  longitudinal and transverse acoustic (LA, TA) phonons, which assist the indirect exciton recombination in BN, as seen in Fig.~\ref{fig:Fig1}(d).
The measured energies are consistent with previous observations~\cite{Vuong2017Dec} and confirm the good control of the isotope purity in the APHT crystals.

To further show the high quality of the monoisotopic BN we build BN/graphene/BN heterostructures in which we use the graphene sheet as a sensitive detector for any disorder within the BN. 
The inset of Fig.~\ref{fig:Fig1}(e) shows such a heterostructure.
To characterize the quality of the encapsulated graphene we employ spatially resolved confocal Raman spectroscopy, see Figs.~\ref{fig:Fig1}(e)~and~(f).
Besides some 'bubbles' within the heterostructure, resulting from a "self-cleaning" mechanism of interfacially trapped hydrocarbons~\cite{Kretinin2014}, we observe a homogeneous Raman response of pristine graphene.
In particular, the width of the 2D-peak is approximately $\Gamma_\mathrm{2D}\approx 16\,\mathrm{cm^{-1}}$ showing very little broadening due to strain variations \cite{Neumann2015b}.
As strain variations are the dominating scattering mechanism in high-quality graphene devices this very low $\Gamma_\mathrm{2D}$ indicates that these heterostructures promise high charge carrier mobilities~\cite{Couto2014}.

The suitability of the monoisotopic BN as substrate for high-performance graphene devices is examined with BN/graphene/BN Hall bar devices fabricated via electron beam lithography and reactive ion etching, see inset of Fig.~\ref{fig:Fig2}(b).
In total we build one device from \textsuperscript{10}BN and five devices from \textsuperscript{11}BN, which all show similar electronic behavior.
The BN thicknesses used in our devices ranges from ${\sim}20\,\mathrm{nm}$ to ${\sim}90\,\mathrm{nm}$. 
First, we characterize the electronic quality of the heterostructure at low temperatures via 4-point electrical transport measurements.
In Fig.~\ref{fig:Fig2}(a) the conductivity $\sigma$ and resistivity $\rho$ at 1.7~K of the heterostructure built with \textsuperscript{10}BN is shown.
The charge neutrality point (CNP) indicated by a sharp increase of the resistance is very close to $V_{BG}=0\,V$ demonstrating low doping and low charge carrier inhomogeneity.  
To quantify the charge carrier density inhomogeneity $n^*$, we employ a double logarithmic plot of $\sigma$ against the charge carrier density $n$, see Fig.~\ref{fig:Fig2}(b).
$n^*$ is defined as the crossing point between two linear fits corresponding to the saturation $\sigma_{min}$ and $\sigma(n)$ at high $n$.
We find low inhomogeneities of $n^*\approx6\times10^9\,\mathrm{cm^{-2}}$; a low value for BN-graphene heterostructures without graphite gates \cite{Couto2014,Banszerus2015,Kretinin2014,Purdie2018}.    

To further probe that the monoisotopic BN does not disturb sensitive quantum states we investigate the quantum Hall effect in our \textsuperscript{10}BN-heterostructure.
Thanks to low $n^*$ and a high carrier mobility of around $\mu \approx 500.000\,\mathrm{cm^2/(Vs)}$, the integer quantum Hall effect emerges at values below 1~T, see Fig.~\ref{fig:Fig2}(c), while four-fold degeneracy lifting \cite{Young2012a} already occurs at $\approx 2$~T, showing that electron-electron interactions are not masked by disorder.
At higher magnetic fields additional quantum Hall plateaus occur at fractional filling factors.
As can been seen in Fig.~\ref{fig:Fig2}(d), we observe fully quantized fractional plateaus at fillings of $2/3$, $4/3$, $5/3$... at magnetic fields of 13.8~T in both the longitudinal resistivity $\rho_{xx}$ and transversal conductivity $\sigma_{xy}$.
The $\nu=1/3$ plateau is not clearly visible in  $\sigma_{xy}$, possibly because of its proximity to the isolating $\nu=0$ state~\cite{Dean2011} and the constant bias voltage measurement technique employed here. 
The well developed fractional quantum Hall states clearly show that the monoisotopic BN is a suitable substrate to investigate fragile quantum states within graphene and possibly in other 2D materials.

\begin{figure}[tb]
	\centering
	\includegraphics{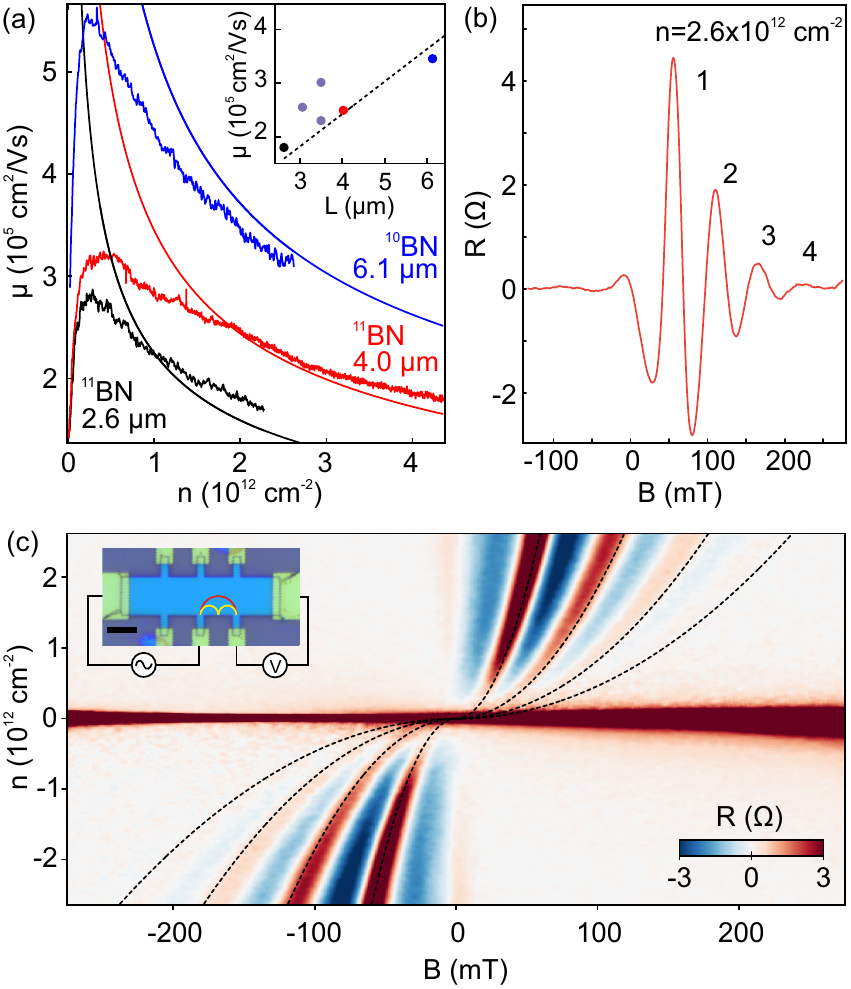}
	\caption{(a) Low temperature mobility as a function of $n$ for different devices. The solid lines indicate the expected mobility values for ballistic transport, see text for details. The inset shows the mobilities at fixed $n=2\times10^{12}\,\mathrm{cm^{-2}}$ for all 6 devices.
	(b) Magnetic focusing of ballistic electrons. Non-local resistance as a function of magnetic field at $n=2.8\times10^{12}\,\mathrm{cm^{-2}}$ shows 4 resonances indicating magnetic focusing. The inset in (c) shows the used \textsuperscript{10}BN/graphene device and the measurement scheme. Scale bar: $5\,\mathrm{\mu m}$.
	(c) Non-local resistance as a function of magnetic field and charge carrier density. The black dashed lines show where the resonance condition $2\nu r_c=L$ is met.   
	}
	\label{fig:Fig3}
\end{figure} 

\begin{figure*}[tb]
	\centering
	\includegraphics{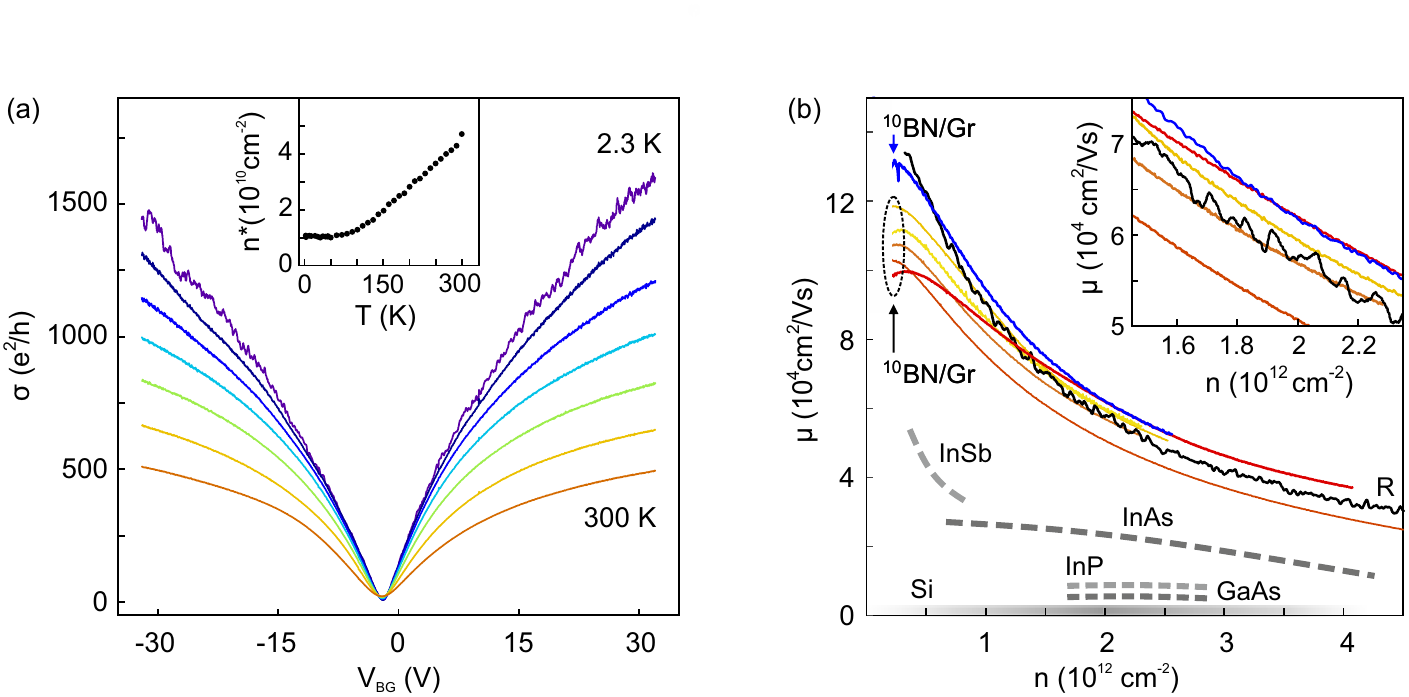}
	\caption{(a) Conductivity $\sigma$ of a \textsuperscript{11}BN device as a function of charge carrier density $n$ for different temperatures. The temperature increment between the data sets is $\approx 50\,\mathrm{K}$. The inset shows the increase of $n^*$ due to thermally excited charge carriers.
    (b) Room temperature mobility as a function of $n$. The blue (red and yellow) curves resemble heterostructures built with \textsuperscript{10}BN (\textsuperscript{11}BN), whereas the black line indicates a reference samples R built with natural BN~\cite{Wang2013}. The inset depicts a close-up of the mobility values at a higher $n$.
	}
	\label{fig:Fig5}
\end{figure*}

In Fig.~\ref{fig:Fig3}(a) we show the charge carrier mobilities $\mu$ of the \textsuperscript{10}BN device and two \textsuperscript{11}BN devices as function of $n$, where the mobility is extracted from $\mu=\sigma/(en)$. 
We find values above $500.000\,\mathrm{cm^2/(Vs)}$, which are only limited by the dimensions of our devices, resulting in quasi-ballistic transport.
In the case of ballistic transport the mean free path $l_m=\sigma h/(2e\sqrt{\pi n})$ is only restricted by the device dimension $L$, thus $l_m\approx L$.
For our samples $L$ represents both the width of the Hall~bar and the distance between two voltage probes.
Considering $\mu=\sigma/(en)$, this implies that for ballistic transport $\mu$ decreases as:  $\mu\approx4eL/\sqrt{\pi h^2 n}$.
As seen in Fig.~\ref{fig:Fig3}(a), we indeed observe a significant decrease of $\mu$ with increasing $n$ and the measured mobilities for higher charge carrier densities are very close or even slightly above this limit.
Similarly, the mobilities increase with the device dimension $L$, see inset in Fig.~\ref{fig:Fig3}(a).
This suggests that the dimensions of the device is limiting the mobility in our devices rather than intrinsic material properties of the graphene or the BN-substrate, indicating ballistic charge carriers. 
To further prove the ballistic nature of the electron transport in our devices we perform magnetic focusing experiments.
To this end we apply a current $I$ to one contact pair, while measuring the non-local voltage $V$ at another pair, see the scheme illustrated in the inset of Fig.~\ref{fig:Fig3}(c).
The applied perpendicular magnetic field B bends the injected charge carriers.
The electrons are focused into the voltage probe, when the cyclotron radius $r_c=\hbar\sqrt{\pi n}/(eB)$ is half of the distance $L$ between the current injector and voltage probe, resulting in a maximum in $R=V/I$.
Crucially, this effect breaks down if there is any scattering of the charge carriers.
Additional maxima occur when the charge carriers are reflected specularly from the sample edge and the resonance condition $2\nu r_c=L$, with $\nu=1,2,3...$, is met.
As shown in Fig.~\ref{fig:Fig3}(c), we observe magnetic focusing resonances starting from charge carrier densities of $|n|<0.25\times10^{12}\mathrm{cm^{-2}}$ and up to an order of 4. 
This shows that the charge carriers travel ballistically over distances of at least $\pi L/2\approx9.6\,\mathrm{\mu m}$, proving that the APHT monoisotopic BN does not introduce any significant scattering and help to maintain the pristine properties of the graphene.
   	
After having shown low-temperature transport data on graphene encapsulated in monoisotopic BN, highlighting device quality very much comparable to state-of-the-art BN/graphene sandwich devices built with HPHT BN, we will now focus on transport measurements at elevated temperatures. 
In Fig.~\ref{fig:Fig5} we show the conductivity of a \textsuperscript{11}BN device as a function of gate voltage (i.e. carrier density) for various temperatures.
While the conductivity decreases due to increased electron-phonon scattering, conductivity values over $500\,e^2/h$ ($\rho_{min}\approx 50\,\Omega$) can still be reached at room temperature with monoisotopic BN/graphene devices.
Importantly, the position of the CNP does not change with temperature, showing that no defects within the BN are thermally excited, which would result in a finite doping of the graphene flake.
Similarly, the charge carrier inhomogeneity, expressed by $n^*$, stays low over the entire temperature range, see the inset of Fig.~\ref{fig:Fig5}(a), only showing the expected broadening due to thermally excited charge carriers within graphene.
In Fig.~\ref{fig:Fig5}(b) we compare the extracted room temperature mobility of our monoisotopic BN/graphene Hall bars to a reference BN/graphene-heterostructure device R based on natural BN~\cite{Wang2013} and to other materials (see thick dashed lines). 
Notably, the shown reference sample R (see black line) is only limited by the intrinsic scattering of electrons with acoustic and optical phonons~\cite{Wang2013,Sohier2017Aug}.
Interestingly, all our fabricated devices are very close to this value and some of our devices (e.g. the one based on \textsuperscript{10}BN/graphene; see blue line) are even outperforming the one made from the natural BN resulting in carrier mobility values being consistently higher for increasing $n$ (see inset in Fig.~4(b)). 
This is surprising and on one hand demonstrates the  high potential of monoisotopic BN for graphene-based applications.
On the other hand this also asks for further investigations to shed more light on the differences of these encapsulation and substrate materials.

To conclude, our work demonstrates excellent transport performance of van der Waals heterostructures based on graphene and atmospheric pressure high temperature grown monoisotopic boron nitride~\cite{Liu2017,Liu2018}. 
Raman spectroscopy and cathodoluminescence spectra indicate that the quality of the APHT monoisotopic BN is very close to the conventional HPHT grown BN.
Our transport measurements show that the charge carrier mobility in our BN/graphene devices is only limited by devices dimensions or temperature, rather than material imperfections.
This shows, that APHT grown BN is a true alternative to HPHT BN.
Since the method of growth is more suitable for extra-large devices and larger scales~\cite{Liu2017,Kumaravadivel2019Jul}, our findings are certainly highly interesting not only for fundamental research, but also for applications based on graphene and related 2D materials.
Importantly, the APHT method also allows for the control of the isotopic concentrations.  
Thus, we believe that the presented work will trigger more studies exploring the potential and benefits of isotopically purified BN substrates.

\begin{acknowledgments}
\textbf{Aknowledgements:}
The authors thank L.~Banszerus, B.~Beschoten and F.~Haupt for helpful discussions.
This project has received funding from the European Union’s Horizon 2020 research and innovation programme under grant agreement No 785219 (Graphene Flagship), and support by the Helmholtz Nanoelectronic Facility (HNF)~\cite{Albrecht2017} at the Forschungszentrum J\"ulich.
\end{acknowledgments}

\bibliography{IsohBN}

\end{document}